# Computationally repurposed drugs and natural products against RNA dependent RNA polymerase as potential COVID-19 therapies

Sakshi Piplani[1-2], Puneet Singh[2], David A. Winkler[3-6], Nikolai Petrovsky[1-2]

1 College of Medicine and Public Health, Flinders University, Bedford Park 5046, Australia

2 Vaxine Pty Ltd, 11 Walkley Avenue, Warradale 5046, Australia

3 La Trobe University, Kingsbury Drive, Bundoora 3042, Australia

4 Monash Institute of Pharmaceutical Sciences, Monash University, Parkville 3052, Australia

5 School of Pharmacy, University of Nottingham, Nottingham NG7 2RD. UK

6 CSIRO Data61, Pullenvale 4069, Australia

**Abstract**

For fast development of COVID-19, it is only feasible to use drugs (off label use) or approved natural products that are already registered or been assessed for safety in previous human trials. These agents can be quickly assessed in COVID-19 patients, as their safety and pharmacokinetics should already be well understood. Computational methods offer promise for rapidly screening such products for potential SARS-CoV-2 activity by predicting and ranking the affinities of these compounds for specific virus protein targets. The RNA-dependent RNA polymerase (RdRP) is a promising target for SARS-CoV-2 drug development given it has no human homologs making RdRP inhibitors potentially safer, with fewer off-target effects that drugs targeting other viral proteins. We combined robust Vina docking on RdRP with molecular dynamic (MD) simulation of the top 80 identified drug candidates to yield a list of the most promising RdRP inhibitors. Literature reviews revealed that many of the predicted inhibitors had been shown to have activity in in vitro assays or had been predicted by other groups to have activity. The novel hits revealed by our screen can now be conveniently tested for activity in RdRP inhibition assays and if conformed testing for antiviral activity invitro before being tested in human trials.

**Introduction**

Since December 2019 and the outbreak of a pneumonic disease called COVID-19 caused by a novel coronavirus (SARS-CoV-2) in China, clinicians globally have faced unprecedented challenges in managing the disease caused by this pandemic virus. There has been a massive international research effort to discover effective drugs and vaccines for this and other pathogenic coronaviruses such as SARS and MERS.[1-17] Design of potent new coronavirus drugs is very important for future disease preparedness. However, for COVID-19, it is only feasible to use drugs that are already registered (off label use) or approved natural products that been through at least phase 1 clinical trials for safety assessment. These agents can be used in man quickly, as their safety and pharmacokinetics are well understood. Computational methods offer considerable promise for rapidly screening such drugs for SARS-Cov-2 activity by predicting their affinities for relevant virus protein targets. Recent papers in prominent journals have reported, for example, the application of computational de novo drug design based on the structures of the SARS-Cov-2 protease. [18-20]

The RNA-dependent RNA polymerase (RdRP) is a promising target for SARS-CoV-2 drug development. It has no host cell homologs so selective RdRP inhibitors should be potentially safer, with lower risks of off-target effects. Zhu et al., recently reviewed the biochemical properties of this critical enzyme and cell-based RdRPs assays suitable for high-throughput screening to discover new and repurposed drugs against SARS-CoV-2.[21] Viral RdRP plays a crucial role in the SARS-Cov-2 replicative cycle. Its active site is highly conserved and accessible, making it a promising drug target. Notably, all DNA and RNA viruses employ RdRP proteins for replication and transcription of viral genes and other viral and host factors (Figure 1), suggesting that computational techniques that prove useful for identifying repurposed drugs against COVID-19 RdRP should be broadly applicable to other pathogenic viruses.[21]

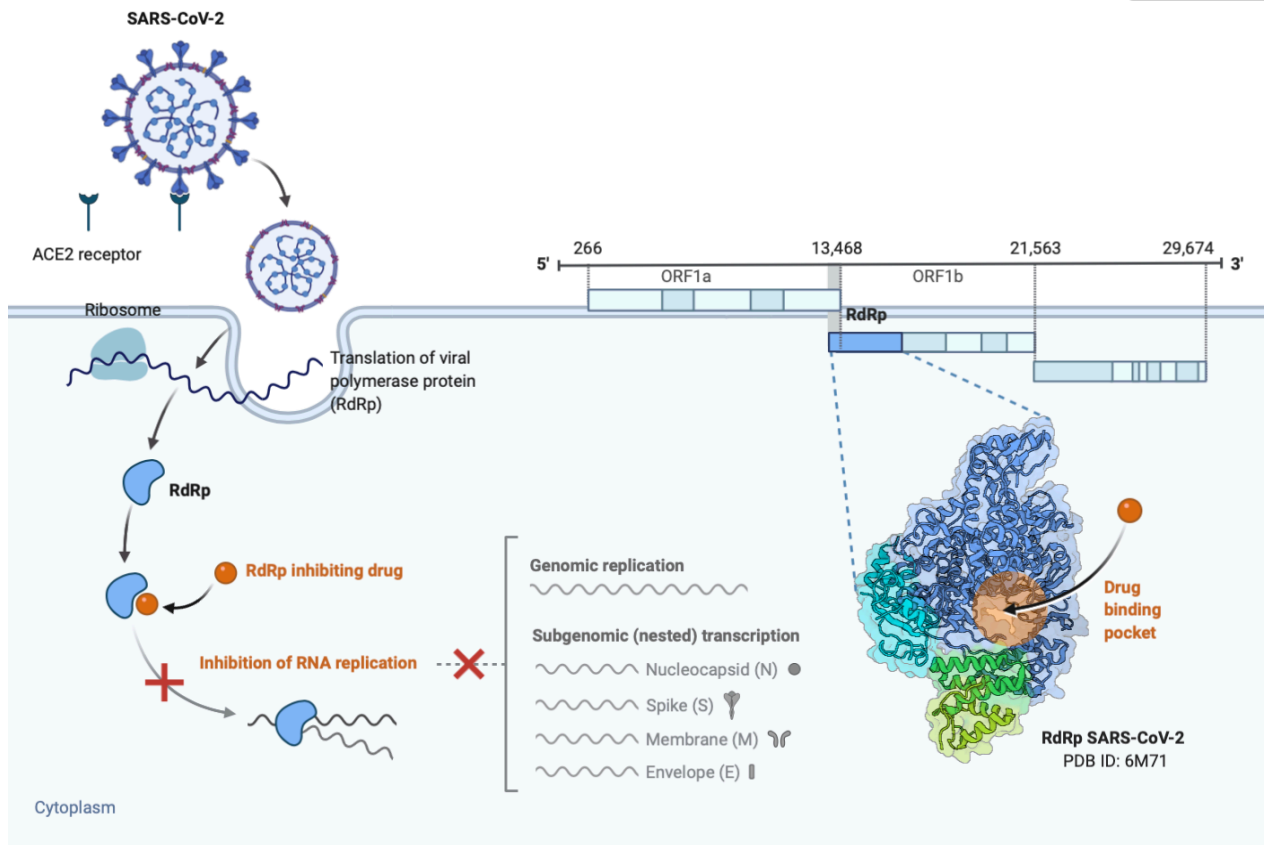

**Figure 1**. Interaction of SARS-Cov-2 with cells showing effects of inhibition of RNA replication by an RdRP inhibitor. Created using BioRender template.

RdRPs share several sequence motifs and tertiary structures amongst all RNA virus types, including positive-sense RNA, negative-sense RNA and dsRNA viruses. The core structure of RdRP resembles a right-hand, with palm, thumb and finger domains. Five of the seven classical catalytic RdRP motifs (A – E) are in the most preserved palm domain, while the remaining two (F and G) are in the finger domains. Five of the seven classic catalytic RdRP (A-E) motifs are in the most highly conserved palm domain, while the other two (F and G) are in finger domain (Figure 2).

The structurally conserved RdRP core and related motifs are important for the catalytic role of viral RdRP and thus represent potential drug targets. While the criteria for the substrates differ, all known RdRPs share the same catalytic mechanism. On host cell infection, viral RdRP participates in formation of the molecular machinery for genome replication by complexing with other factors. It initiates and regulates the elongation of the RNA strand, which involves the addition of hundreds to thousands of nucleotides. When incorporated into the newly synthesised RNA chain, nucleotide analogues such as remdesivir block the RNA elongation catalysed by RdRP (Figure 2).

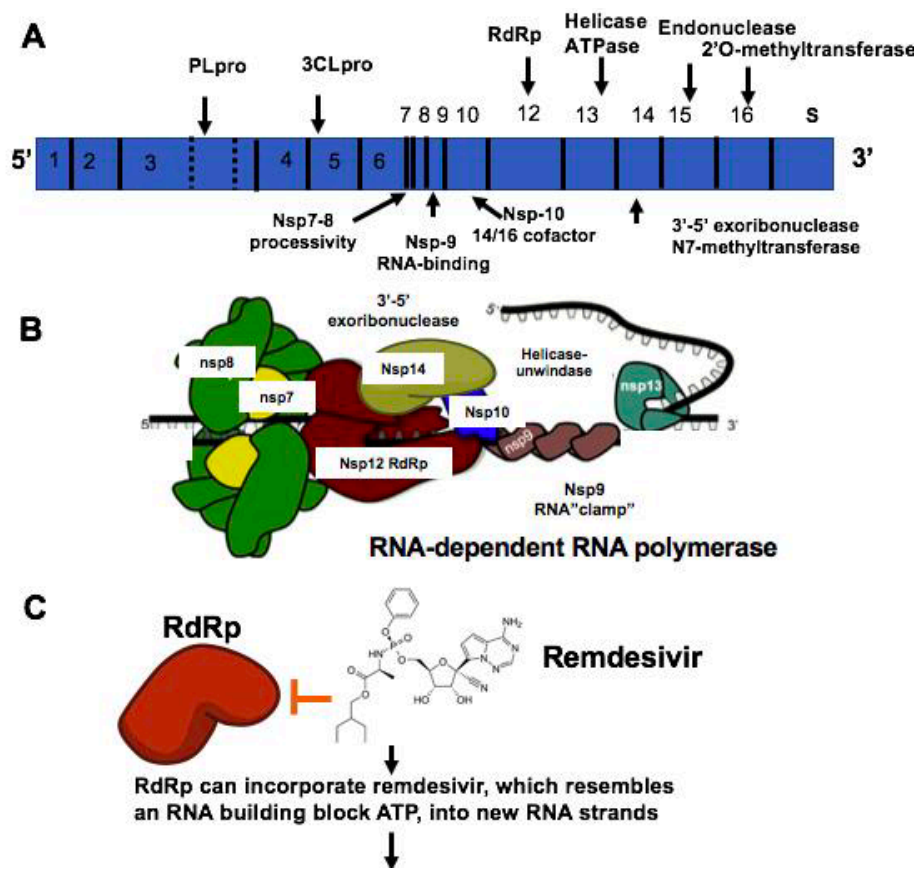

**Figure 2**. Inhibition of RNA-dependent RNA polymerase (RdRp) by repurposed drugs. (A) The SARS-Cov-2 genome composition single strand RNA model. (B) RdRp mediated RNA replication during coronavirus infection. (C) Drugs such as remdesivir inhibit RdRp and block RNA replication. Creative Commons licence figure adapted from Huang et al.[22]

In pandemics, computational methods can quickly identify candidate drugs for repurposing, where speed is of utmost importance. Here we show how a well validated molecular docking followed by a high throughput molecular dynamics simulation (MDS) drug screening program can screen a large number of drugs and natural products to produce a short list of RdRp-inhibiting drug candidates. MDS calculations were used to predict the optimal binding poses and binding energies for 80 of the top hits from virtual screening on SARS-CoV-2 RdRP. Finally we ranked the top candidates for COVID-19 based on binding affinity and novelty.

**Results and discussion**

There have been many computational studies aiming to predict which existing drugs and natural products may be inhibit the main protease, $M^{pro}$, of SARS-Cov-2, but fewer studies have studied drugs for targeting the viral RdRP. The computational workflow used to estimate the relative binding affinities of drugs for the RdRP binding pocket are summarized in Figure 3. The combination of robust Vina docking followed by MDS of the top 80 candidates yields improved performance relative to Vina docking alone.[23] The binding free energies calculated by both computational methods (see Methods) correlated very well ($r^2$=0.84) and the free energies calculated by the thermodynamic cycle correlate with the Vina docking scores with $r^2 = 0.64$. The size of the binding site (area 2920 Å$^2$ and volume 5335 Å$^3$) tends to select larger ligands, many of which are quite flexible. Binding energy penalties due to ligand entropy are likely to be significant. Hence, substantial correlation between the Vina scores and the binding energies from MMBBSA and thermodynamic cycle are important because these algorithms treat ligand entropy approximately and in different ways.[24]

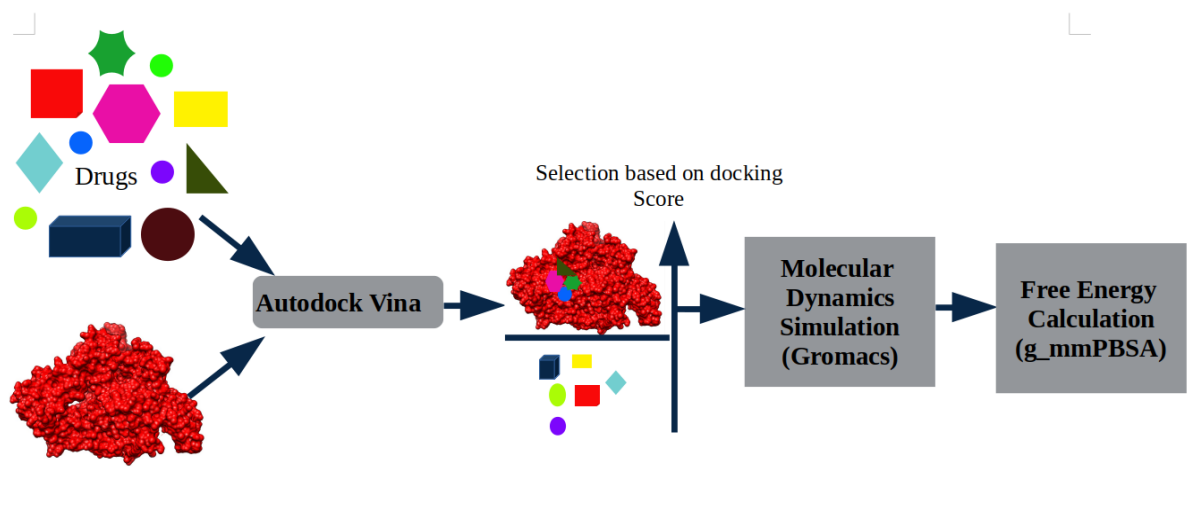

**Figure 3**. Computational workflow for repurposing drugs against SARS-Cov-2 RdRP.

The binding energies of the 80 top ranked ligands from the docking calculations are listed in Supplementary Table 1. The 20 drugs predicted to have the highest binding to RdRp are summarized in Table 1, together with both their MMPBSA and thermodynamic cycle binding energies. The high correlation between the two methods of calculating binding free energies meant that the ranking of the top 20 compounds was very similar when ordered by either binding free energy estimate. Antiviral drugs, Paritaprevir, Beclabuvir, Remdesivir, Voxilaprevir, Setrobuvir, Galidesvir, Elbasvir, Ciluprevir, Faldaprevir, and Tegobuvir account for half of the list. The remaining 10 hit compounds, which include several natural products, are used to treat a diverse range of afflictions – cancers, infections, cardiac insufficiency, liver damage, circulatory issues, and parasitic infections. Almost all of the drugs in the top 20 have relatively large, complex structures and substantial ligand flexibility. Some have had their in vitro activity against SARS-Cov-2 determined experimentally (Figure 4), further supporting our binding predictions.

**Table 1**. Structures and binding energies of 20 top ranked (by MMPBSA score) small molecule ligands to SARS-Cov-2 RDRP.

| Database ID C=ChemBL D=Drugbank | DrugName | Structure | $\Delta G_{MMPBSA}$ ($\Delta G_{thermo}$) kcal/mol |
|---|---|---|---|
| C 3391662 | Paritaprevir (antiviral) | 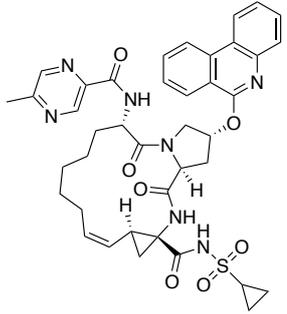 | -54.3 (-67.5) |
| C 1200633 | Ivermectin (anti-parasitic) | 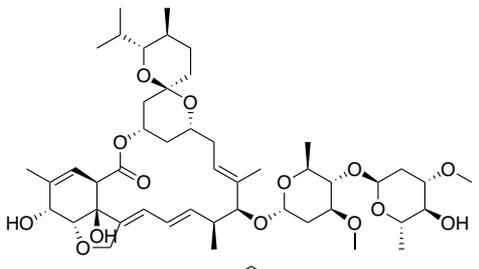 | -54.1 (-69.7) |
| C 3126842 | Beclabuvir (antiviral) | 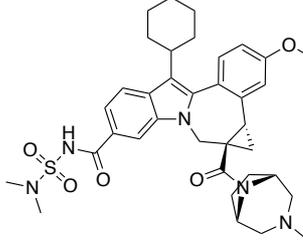 | -53.5 (-66.4) |
| C 3809489 | Bemcentinib (anticancer) | 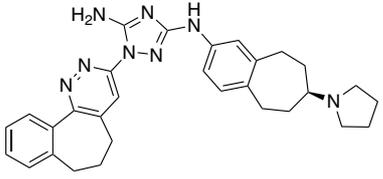 | -46.1 (-62.5) |
| D 14761 | Remdesivir (antiviral) | 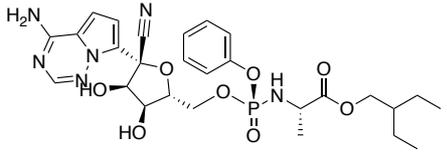 | -44.6 (-56.8) |
| C 1751 | Digoxin (cardiac drug) | 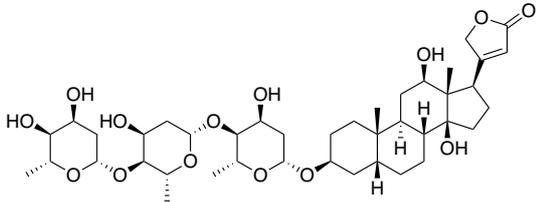 | -41.2 (-54.4) |
| D 09298 | Silibinin (chemo-protectant | 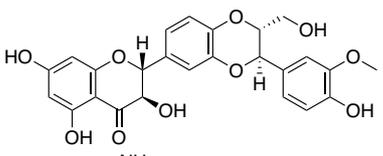 | -40.2 (-57.2) |
| C 1236524 | Galidesvir (antiviral) | 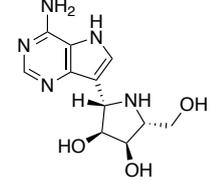 | -40.2 (-49.2) |

| Database ID C=ChemBL D=Drugbank | DrugName | Structure | ΔG$_{MMPBSA}$ (ΔG$_{thermo}$) kcal/mol |
|---|---|---|---|
| C 1076263 | Setrobuvir (antiviral) | 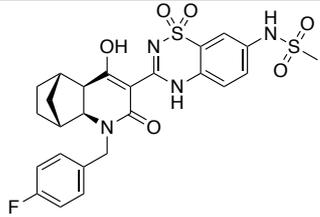 | -40.0 (-51.1) |
| C 3707372 | Voxilaprevir (antiviral) | 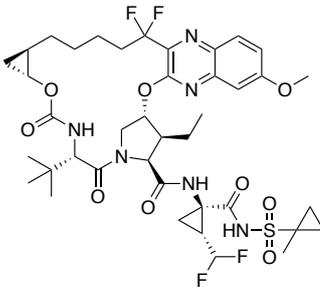 | -39.5 (-52.8) |
| C 2013174 | Vedroprevir (antiviral) | 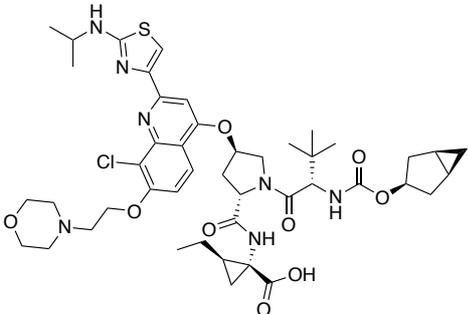 | -38.4 (-44.6) |
| C 1241348 | Faldaprevir (antiviral) | 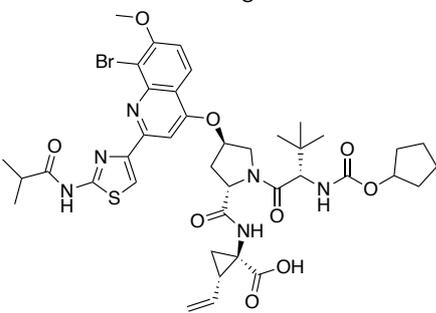 | -38.1 (-46.4) |
| C 413 | Sirolimus (Rapamycin) (immuno-suppressant) | 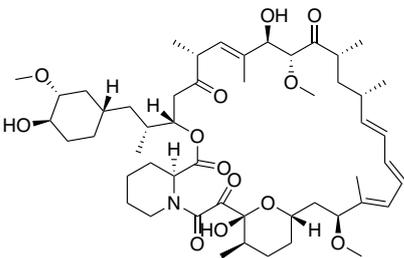 | -37.4 (-45.2) |

| Database ID C=ChemBL D=Drugbank | DrugName | Structure | ΔG$_{MMPBSA}$ (ΔG$_{thermo}$) kcal/mol |
|---|---|---|---|
| C 3301668 | Carbetocin (anti-hemorrhagic) | 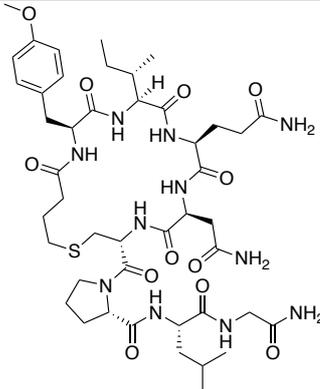 | -37.4 (-35.1) |
| D 01051 | Novobiocin (antibiotic) | 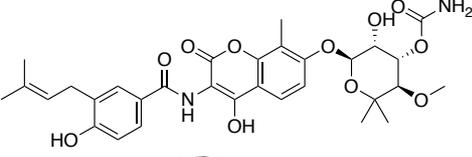 | -37.3 (-45.3) |
| C 1683590 | Eribulin (anticancer) | 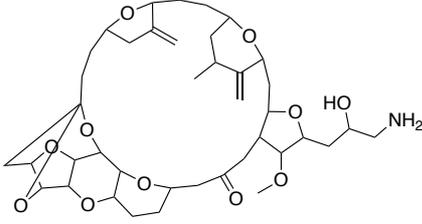 | -36.5 (-48.3) |
| C 442 | Ergotamine (anti-migraine) | 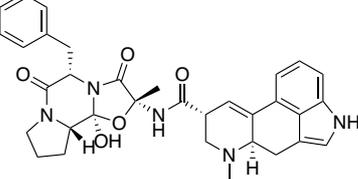 | -36.2 (-45.6) |
| C 1957287 | Tegobuvir (antiviral) | 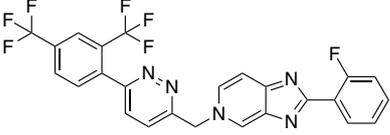 | -34.7 (-46.0) |
| D 12466 | Favipiravir (antiviral) | 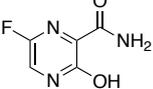 | -34.2 (-41.6) |
| D 14850 | Deleobuvir (antiviral) | 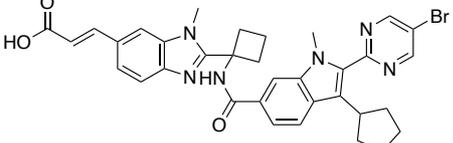 | -34.1 (-39.6) |

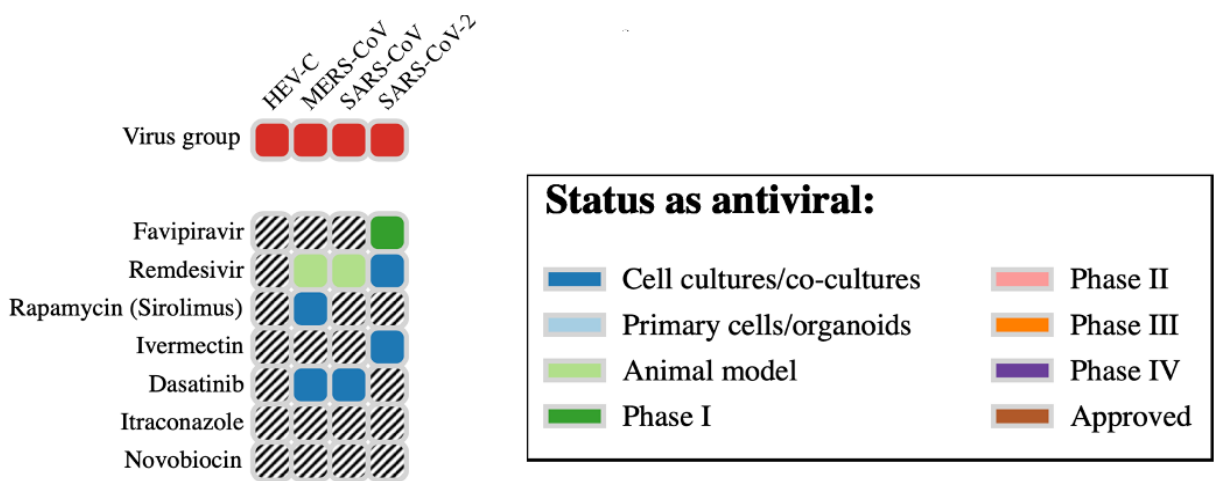

**Figure 4**. Experimental antiviral spectrum for hit compounds. From DrugVirus.info.

*Antiviral agents*.

The calculated binding energy of several of the antiviral drugs, Paritaprevir and Beclabuvir, are very similar, within the calculation uncertainties in energies. Several of the top antiviral agents predicted by this study, have also been identified in other *in silico* docking studies of small molecule drugs. This provides a degree of validation that our computational methods are appropriate and are yielding similar results to the other published studies for some well-studied antiviral drugs.

Beg and Athar used a homology modelled structure for SARS-Cov-2 RdRP and AutoDock Vina to rank the binding of a range of antiviral drugs to the RdRP enzyme.[25] Paritaprevir, Beclabuvir, and Favipiravir had some of the highest docking scores. Cozac et al. reported a combined docking and machine learning study of inhibitors of RdRP from HCV, poliovirus, dengue virus, and influenza virus.[26] They identified Faldaprevir, Vedroprevir, Beclabuvir and Remdesivir as having good docking scores to SARS-Cov-2 RdRP and predicted viral RdRPs $IC_{50}/EC_{50}$ values below 5μM by multiple machine learning classification models. A homology model for SARS-Cov-2 RdRP and AutoDock was used by Dutta et al. to rank antiviral drugs for potential use in treating COVID-19 patients.[27] One of the most promising drugs for repurposing was Beclabuvir with a predicted docking score of (-10 kcal/mol) and IC$_{50}$ of 50nM, although several other antiviral agents had similar docking scores, e.g. tegobuvir (-9.7 kcal/mol), dasabuvir (-9.4 kcal/mol), lomibuvir (-11 kcal/mol), setrobuvir ( -10.5 kcal/mol) and dasabuvir (-10.4 kcal/mol). Remdesivir (-7.4 kcal/mol), radalbuvir (-7.4 kcal/mol) and dasabuvir (-6.7 kcal/mol) had significantly lower docking scores. None of these studies used MDS refinement of docked compounds to improve the estimates of the binding free energy of the drugs.

Remdesivir has been predicted to be active against SARS-Cov-2 RdRP in several studies.[28-30] There is limited *in vitro* evidence for activity of remdesivir with an IC$_{50}$ 3.7 μM in Vero cells,[31, 32] and recent clinical studies suggest its efficacy in treating COVID-19 infection in man is limted.[33]

Elfiki reported two studies aimed at repurposing existing drugs against RdRP.[29, 30] In one study, AutoDock Vina docking, and MDS (50 ns production runs) was used to model viral RdRp and calculate the binding affinity of several drugs and drug candidates. Existing drugs, Sofosbuvir, Ribavirin, Galidesivir, Remdesivir, Favipiravir, Cefuroxime, Tenofovir, Setrobuvir, and Hydroxychloroquine, were all predicted to bind to RdRp. Setrobuvir, YAK, and IDX-184 had even more favourable binding energies, and four novel derivatives of the latter drug showed binding to SARS-CoV-2 RdRp. The second study used only docking to predict the binding of the same drugs to RdRP from SARS-Cov-2, SARS, and HCV. HCV RdRp had greater binding to the antiviral drugs than did SARS-Cov-2 RdRP, with SARS RdRP having the lowest affinity as ranked by Vina docking scores.

Ahmed et al. used molecular docking and 100 ns MDS to estimate the relative binding affinities, interactions, and structure-activity-relationships of 76 prescription antiviral drugs for RdRp.[34] MDS on the best docking candidates showed that remdesivir, raltegravir, and simeprevir had the best binding free energies, ranging from –32 to –38 kcal.mole. Aouldate et al. reported virtual screening of 50,000 chemical compounds from the CAS Antiviral COVID19 database against RdRP.[35] They used a combination of 3D-similarity searches, docking, and 20 ns MDS to rank and select lead compounds and reported one compound (833463-19-7) that bound well to RdRP.

Banerjee et al. used protein modelling and computational docking techniques to investigate the effects of common mutations in RdRp, 3CLpro and PLpro sequences of Indian patients.[36] Two RdRp mutations occurred in the Indian population with prevalence >5% and these were analysed as possible targets for repurposed drugs. Docking using Autodock Vina predicted Elbasvir as the best inhibitor of RdRp in the Indian population, followed by Remdesivir and Methylprednisolone.

*Natural products and analogues.*

Given the large degree of interest in repurposing antiviral drugs for use in treating COVID-19, natural product hits are of greater interest given their relative novelty. The drug with the strongest binding affinity to RdRP in our study was the natural anti-helminthic product, Ivermectin. Ivermectin and other avermectins and milbemycins are broad spectrum antiparasitic macrocyclic lactones derived from the bacterium *Streptomyces avermitilis*. Ivermectin's mode of action is by enhancing inhibitory neurotransmission by binding to glutamate-gated chloride channels. Ivermectin has been shown to be effective against several positive-sense single-strand RNA viruses,[37] including SARS-CoV-2,[38, 39] and has been suggested by others as a COVID-19 drug repurposing candidate.[40] It has demonstrated broad in vitro antiviral activity, including against positive-sense single-strand RNA viruses such as SARS-CoV-2.[37] It inhibited replication of SARS-CoV-2 in monkey kidney cell culture with an $IC_{50}$ of 2.2 - 2.8 µM.[32, 37, 41] It has been predicted to inhibit SARS-Cov-2 RdRP in several computational studies.[42, 43] Parvez et al. used MDS techniques, augmented by very short

MDS to predict the binding affinities of Ivermectin and several other antiviral agents included in Table 1 to RdRP.[43] Janabi et al. also predicted favourable binding energies for Ivermectin and several milbemycins to RdRp using AutoDock Vina, but without subsequent simulation of the protein-ligand complexes.[42] Figure 5 summarizes the main interactions between Ivermectin and the binding site of RdRP and the docking pose from the MDS.

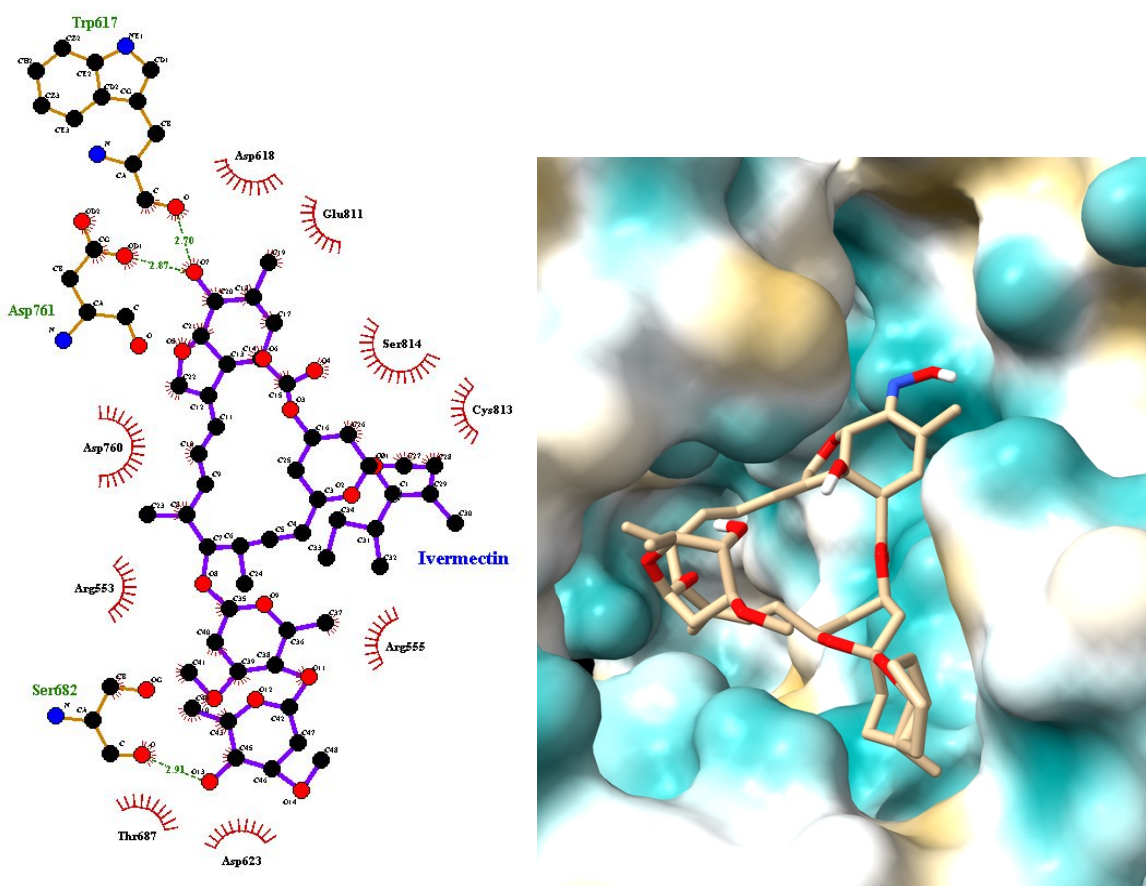

**Figure 5**. LigPlot (left) and hydrophobic protein surface representation (right) of the main interactions between RdRP and ivermectin.

Kalhor et al predicted digoxin would inhibit the interaction of the RBD domain of the SARS-CoV-2 with the ACE2 receptor by but no computation studies other than ours have reported the possibility of RdRP inhibition by digoxin. However, very recently digoxin was shown to have potent *in vitro* antiviral effects against SARS-Cov-2 in Vero cells ($IC_{50}$ 37 nM) and by virus growth kinetics[44] with an $IC_{50}$ of 190nM in Vero cells.[45] The activity of digoxin in these assays was substantially better than that of chloroquine or remdesivir. Figure 6 shows the main interactions of digoxin with the RdRP binding site and how the drug binds to the catalytic cavity. The main interactions are also listed in Supplementary Table 2.

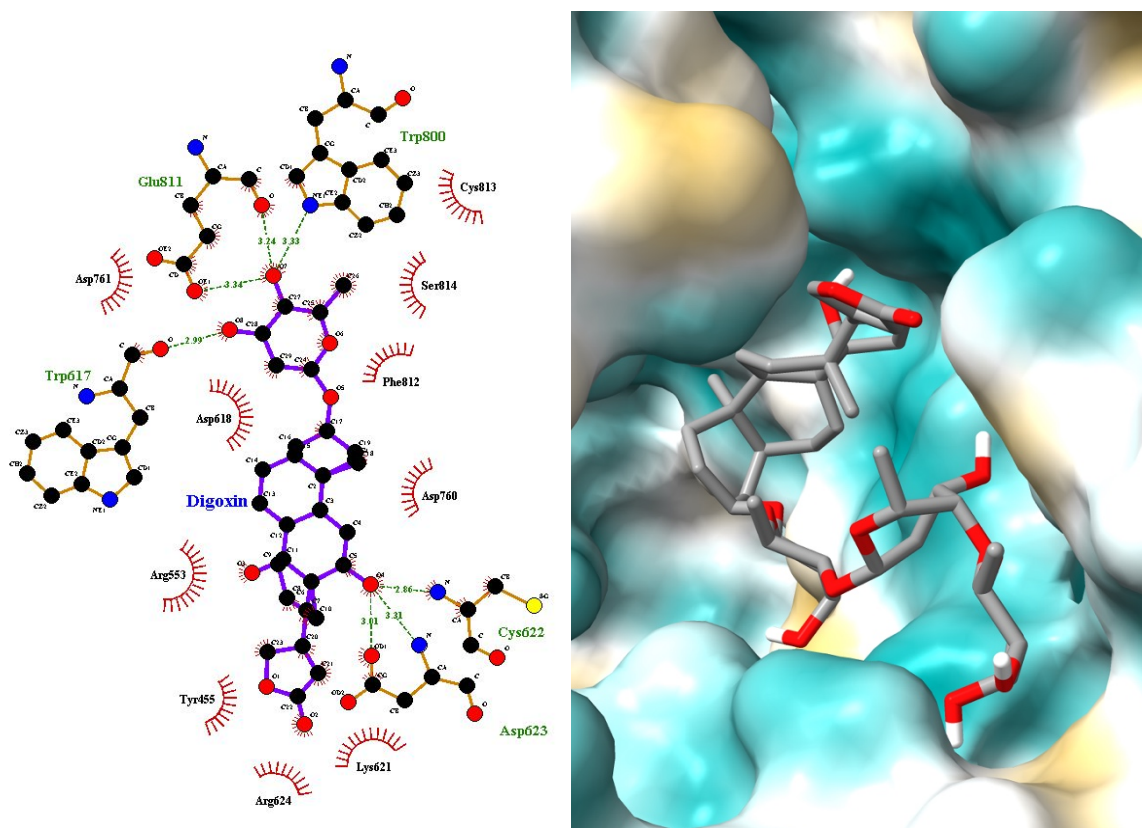

**Figure 6**. LigPlot (left) and hydrophobic protein surface representation (right) of the main interactions between RdRP and digoxin.

Silibinin has been predicted to be a potential inhibitor of SARS-Cov-2 RdRP by two computational docking studies.[46, 47] Bosch-Barrera et al. speculated that Silibinin may have synergistic benefits for treating COVID-19 patients as, apart from its potent antiviral properties against HCV and HIV, it also combated the cytokine storm that causes major problems in COVID-19 patients. Silibinin is the subject of clinical trials planned for the near future. Figure 7 shows the main interactions of Silibinin with the binding site of RdRP and how it binds to the RdRP catalytic cavity. The main interactions are also listed in Supplementary Table 2 for reference.

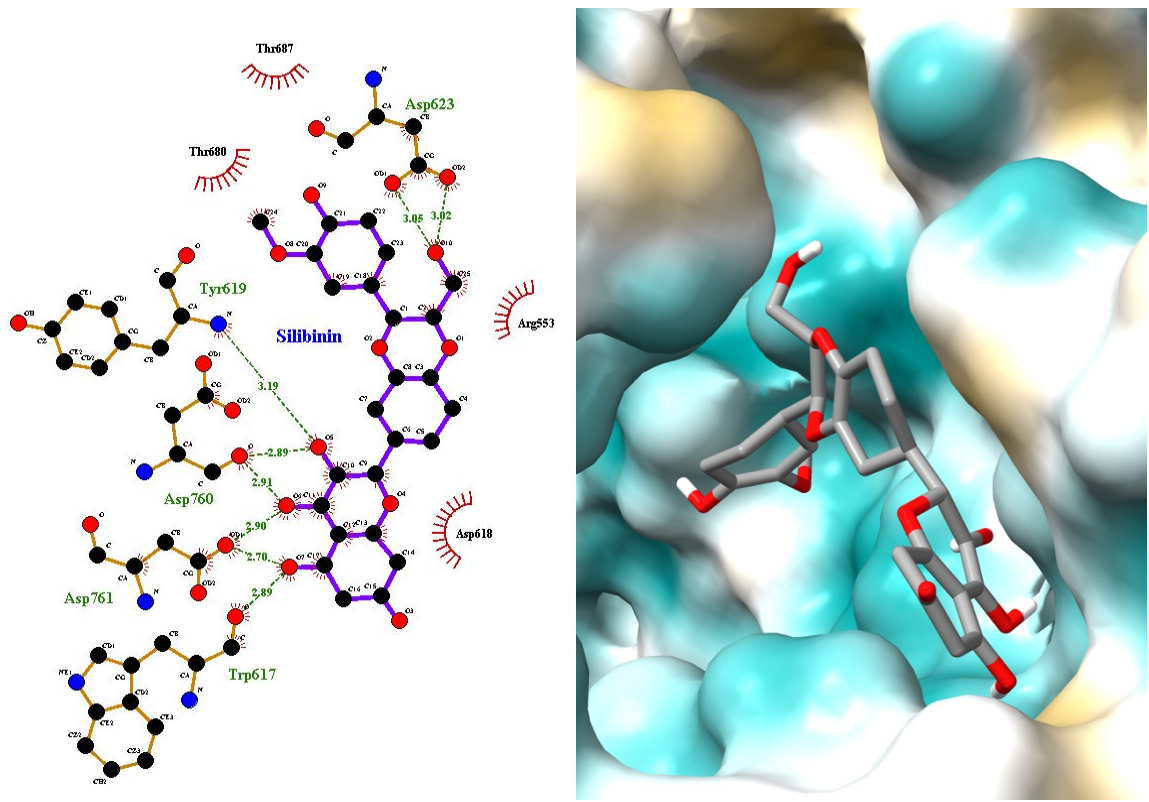

**Figure 7**. LigPlot (left) and hydrophobic protein surface representation (right) of the main interactions between RdRP and silibinin.

AutoDock Vina and MDS using the CHARMM forcefield were used by Pokhrel et al. to predict repurposing of existing drugs and natural products.[48] Sirolimus (Rapamycin) was on the top few hits in their screening and simulation study, suggesting a role for this drug in treating COVID-19 and it is now the subject of a trial of COVID-19 treatment (https://clinicaltrials.gov/ct2/show/NCT04461340). Figure 8 shows the main interactions of sirolimus and the RDRP binding site, and the binding pose from the MDS.

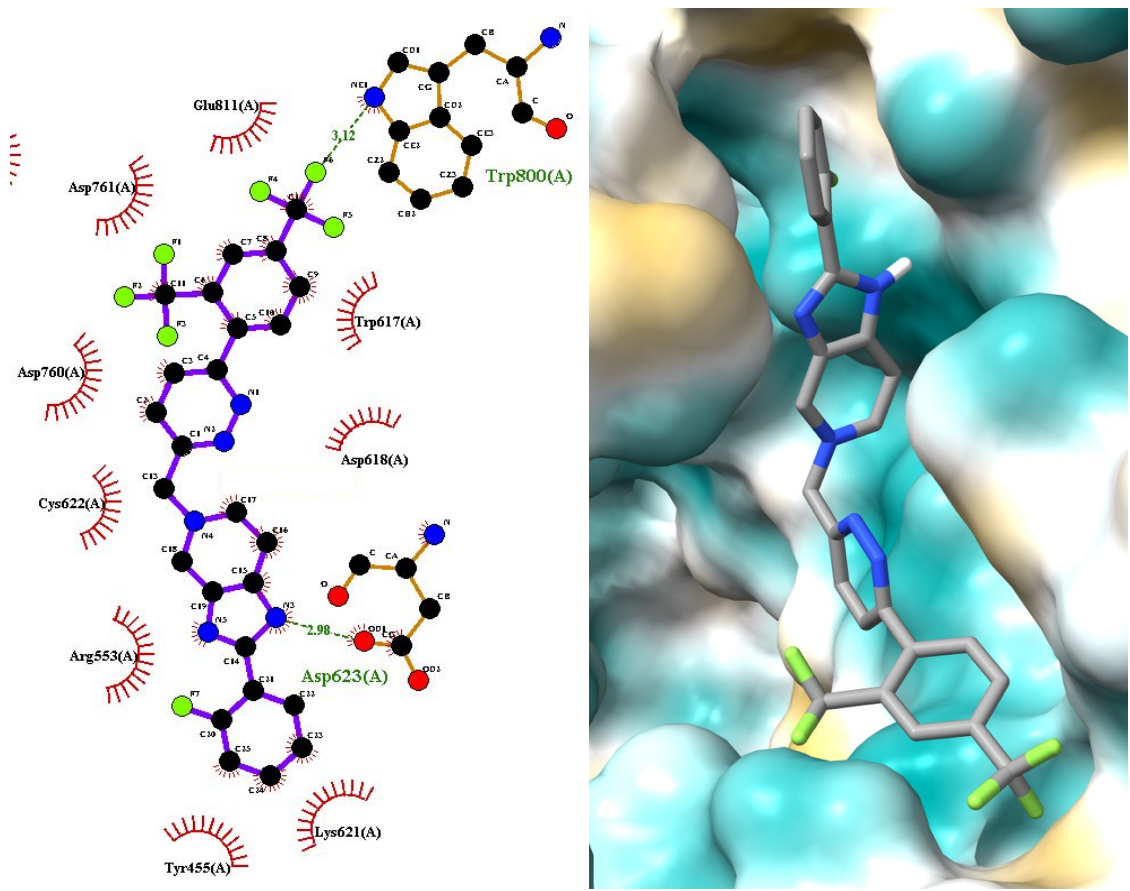

**Figure 8**. LigPlot (left) and hydrophobic protein surface representation (right) of the main interactions between RdRP and Sirolimus (rapamycin).

Carbetocin is a synthetic analogue of the natural hormone, oxytoxin, in which a labile disulfide bond in the macrocycle is replaced by a thioether. Carbetocin was predicted to be one of the top 10 inhibitors of RdRp (single chain) by Ahmad et al.[49] Their studies showed stable binding with a docking score of -9.5 kcal/mol with 8 strong hydrogen bonding interactions with the active site of the enzyme by the Glide docking-scoring function but no MDS was applied to their hits.[49]

Eribulin is a fully synthetic macrocyclic ketone analogue of the marine natural product, Halichondrin B, and is a potent anti-mitotic cancer agent. There are no reports of Eribulin binding to SARS-Cov-2 RdRP but Machitani et al. suggested its COVID-19 use by virtue of its known activity against other viral RdRPs.[50]

Novobiocin was also selected as one of the top five best RdRP binders by Choudhury et al. using the MoleGro virtual docker software.[51].

Ergotamine and related ergot alkaloids have been predicted to be high binders to SARS-Cov-2 molecular targets, including the main protease, M[pro]. [20] There are a few literature reports of its binding to the SARS-Cov-2 RdRP enzyme.[52] There is an in silico predicted SARS-Cov-2 $IC_{50}$ of 190μM.[53] Figure 9 shows the main interactions between the RdRP active site residues and ergotamine, and also illustrates the binding pose from the MDS. The interactions are also listed in Supplementary Table 2.

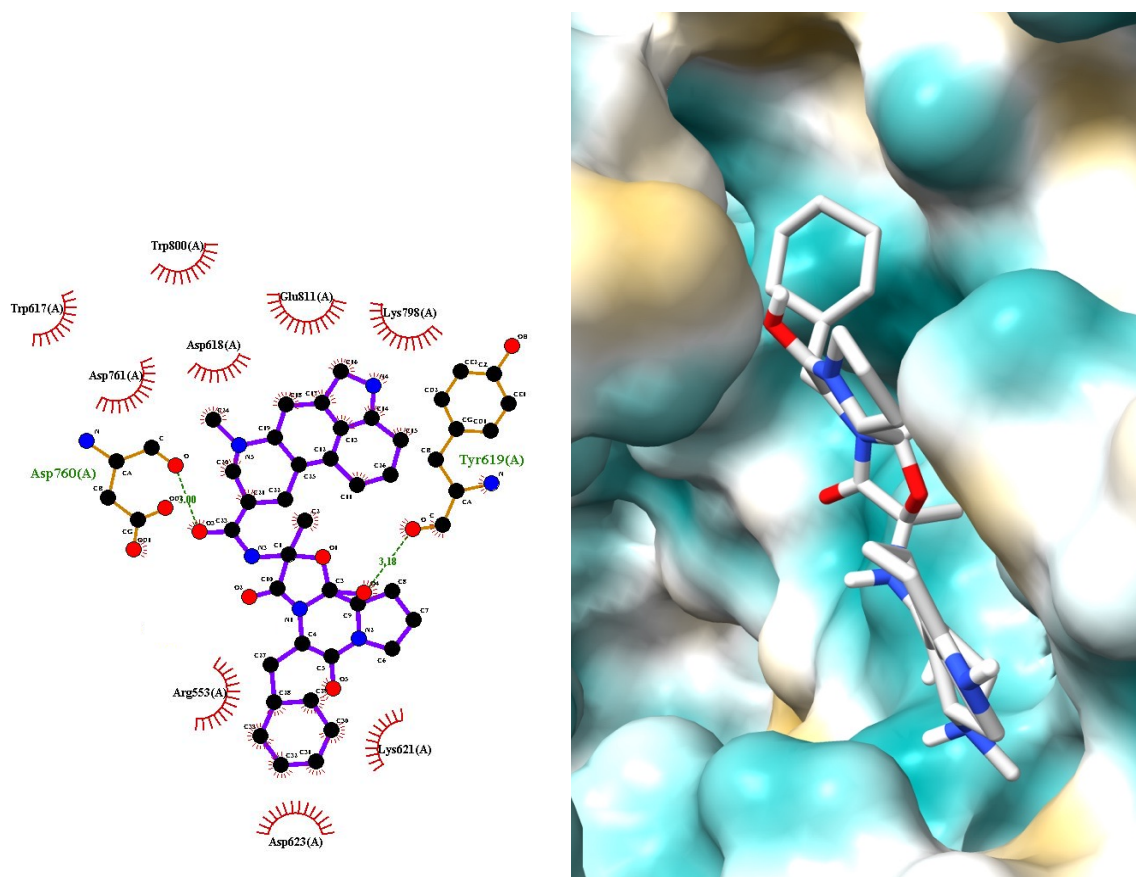

**Figure 9**. LigPlot (left) and hydrophobic protein surface representation (right) of the main interactions between ergotamine and RdRP.

Computational studies on natural products have been reported recently. Singh et al. reported a computational study of the affinity of plant-derived polyphenols as potential inhibitors of

SARS-CoV-2 RdRp.[54] They used AutoDock Vina followed by MDS using Amber to predict the binding poses and affinities of 100 polyphenols. The top four polyphenols had Vina binding energies close to –10 kcal/mol, significantly stronger than the predicted binding of remdesivir.

Wu et al. generated homology models of 18 SARS-Cov-2 viral proteins and two human targets and used structure-based virtual ligand screening to identify potential inhibitors of these targets.[47] They identified a range of drugs and natural products that their computational methods suggested may inhibit the RdRP, including beclabuvir.

*Other drugs*

Bemcentinib selectively inhibits AXL kinase activity, which blocks viral entry and enhances the antiviral type I interferon response. It has been identified as a potential inhibitor of the SARS-Cov-2 M$^{pro}$ enzyme in many computational studies.[20] We have not been able to locate any other published computational publications that suggest that bemcentinib may inhibit the SARS-Cov-2RdRP. Bemcentinib exhibits useful in vitro activity against SARS-Cov-2 with Liu et al. reporting 10-40% protection at 50μM in Vero cells.[55] It was also reported to exhibit an IC$_{50}$ of 100nM and CC$_{50}$ of 4.7μM in human Huh7.5 cells and an IC$_{50}$ of 470nM and CC$_{50}$ of 1.6μM in Vero cells,[56] considerably higher activity than that reported by Liu et al. As a result Bemcentinib is an investigational treatment for COVID-19 (www.clinicaltrialsregister.eu). Figure 10 shows a LigPlot representation of the interactions of key functional groups in bemcentinib with RdRP active site residues, and the binding pose of the drug to the enzyme active site. The interactions are also listed in Supplementary Table 2 for reference.

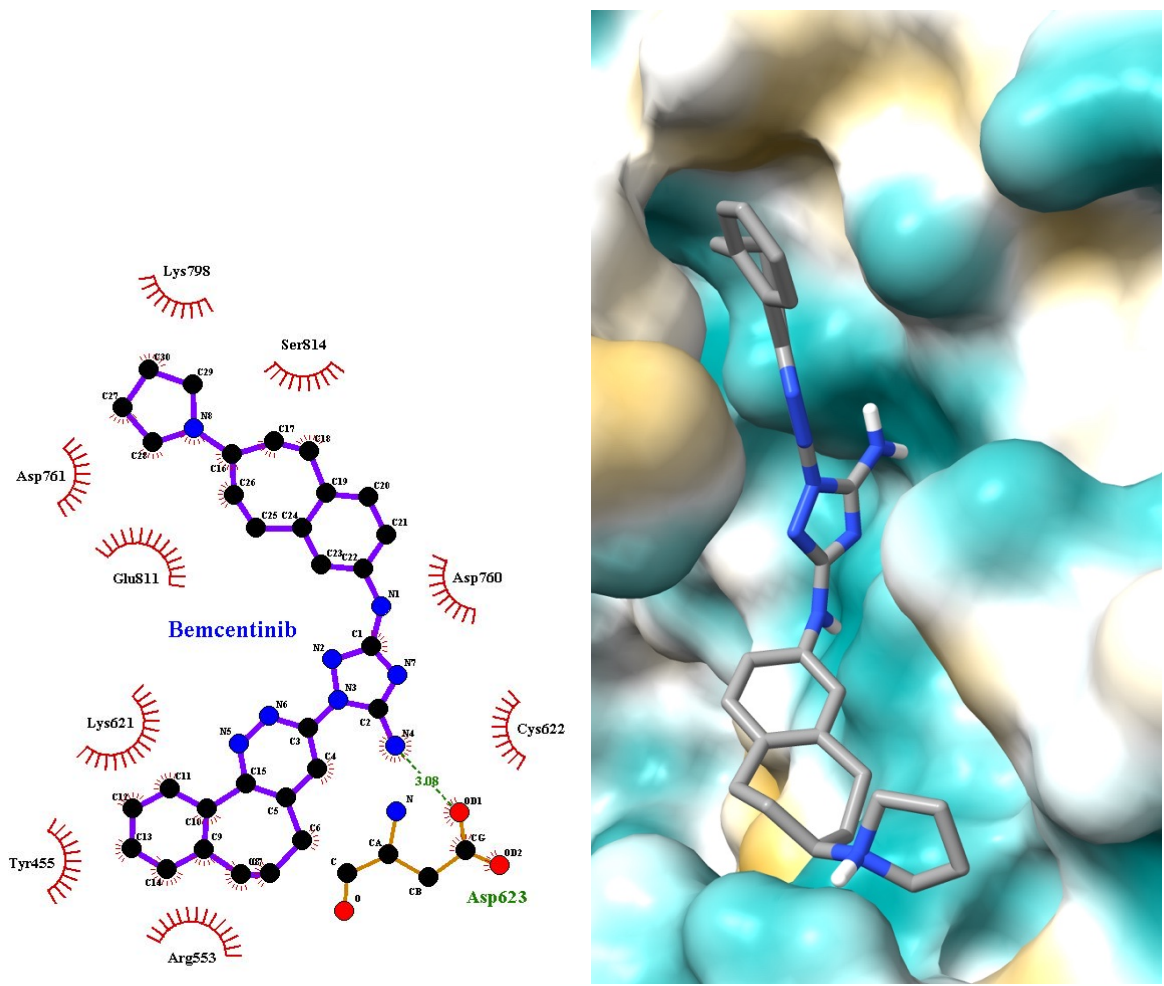

**Figure 10**. LigPlot (left) and hydrophobic protein surface representation (right) of the main interactions between RdRP and bemcentinib.

Computational docking studies identified an unusual arsenic drug, Darinaparsin, as being an active inhibitor of SARS-CoV-2 RdRp.[57] These researchers performed docking calculations on 12 arsenical drugs using iGEMDOCK (http://gemdock.life.nctu.edu.tw/dock/igemdock.php) and AutoDockVina. They screened the arsenical drugs against several other SARS-Cov-2 drugs and additional several promising leads. Neither of these computational studies used MDS to simulate the highest scoring docked structures and calculate more accurate binding energies.

*Other novel putative RdRP inhibitors from the short list of 80 drugs*

Apart from the drugs discussed above, several other drugs in the Supplementary Table 1 are of interest. The top 80 list is strongly populated by antiviral drugs such as ciluprevir, glecaprevir, indinavir, simeprevir, elbasvir and ruzasvir. There are several kinase inhibitors with good predicted binding affinities to RdRP including imatinib, ponatinib, rebastinib, lonafarnib, tivantinib and entrectinib. Antibiotics also feature in the list of the 80 best binders e.g. quinupristin, dalfopristin, rifapentin and erythromycin. Other drugs with binding energies lower than –25 kcal/mol include hesperidin, eltrombopag (also active against Mpro), dutasteride, etoposide, quarfloxin, epirubicin, telmisartan, brequinar, conivaptan, rifapentine, sertindole, itraconazole, vapreotide, bafilomycin A1, bromocriptine, idarubicin, midostaurin and rutin. As Supplementary Table 1 shows, a substantial percentage of the hits generated by our docking and MDS protocols have already been shown to have *in vitro* activity against SARS-Cov-2, or have been predicted to bind to the SARS-CoV-2 RdRP. Interesting some of the compounds have also been predicted to bind several other SARS-Cov-2 protein targets suggesting useful multimodal action that may predict a more potent anti-viral effect. Some identified drugs have already been tested in patients or are undergoing clinical trials in COVID-19. For example, ciluprevir inhibits SARS-Cov-2 Mpro with an $IC_{50}$ of 21 µM,[58] indinavir inhibits SARS-Cov-2 with an $EC_{50}$ >10 µM and $CC_{50}$ >50 µM (A549-hACE2 cells) [59] and an $EC_{50}$ of 59 µM and $CC_{50}$ >81 µM in Vero cells.[60] Simeprevir inhibits SARS-Cov-2 in vitro with an $EC_{50}$ of 4 µM and $CC_{50}$ 19 µM (Vero cells), and exhibits SARS-Cov-2 Mpro inhibition with an $IC_{50}$ = 10 µM.[61] Eltrombopag inhibits SARS-Cov-2 in vitro with an $IC_{50}$ of 8 µM and $CC_{50}$ >50 µM (Vero cells),[62] and an $IC_{50}$ of 8 µM in Vero and Calu-3 cells.[63] Elbasvir also inhibits SARS-Cov-2 in vitro with an $EC_{50}$ of 23 µM in Huh7-hACE2 cells.[64] Telmisartan is in clinical trials for COVID-19 treatment (NCT04356495).[65] Similarly, brequinar has an experimental SARS-Cov-2 $EC_{50}$ of 0.3 µM and $CC_{50}$ > 50 µM in Vero E6 cells,[66] while imatinib produces an 80% reduction in Mpro activity at 10 µM, with an $EC_{50}$ =

8 μM (A549 cells),[67] and an experimental SARS-Cov-2 $IC_{50}$ of 3-5μM, with a $CC_{50}$ >30 μM,[68] Conivaptan also exhibits an $IC_{50}$ of 10 μM against SARS-COV-2 in Vero cells,[69] 4μM in ACE2-A549 cells and a $IC_{50}$ of ~ 10μM against SARS-Cov-2 Mpro in 293T cells.[67] Ponatinib exhibits an experimental SARS-Cov-2 $EC_{50}$ of 1 μM and a $CC_{50}$ of 9 μM in HEK-293T cells,[64] and grazoprevir also shows experimental SARS-CoV-2 inhibition, with $EC_{50}$ of 16 μM and $CC_{50}$ of >100 μM in Vero E6 cells.[64] Finally, itraconazole has an experimental SARS-Cov-2 Mpro $IC_{50}$ of 110 μM[70] and SARS-Cov-2 $EC_{50}$ = 2.3 μM in human Caco-2 cells,[71] idarubicin shows weak in vitro Mpro activity with an $IC_{50}$ of 250−600μM,[72] and ciclesonide shows an in vitro $EC_{90}$ for SARS-CoV-2 of 5μM in Vero cells, 0.55μM in differentiated human bronchial tracheal epithelial cells, blocks viral RNA replication, and supresses replication of 15 mutants by >90%[62,73] It is used to treat COVID-19 patients.[74]

As was demonstrated when we applied the same protocol to repurposing of drugs and natural products acting against the SARS-Cov-2 main protease, Mpro, our computational methods generate short targeted lists of existing drugs that may be useful for treating COVID-19.[20]s As well as direct antiviral activities, some of the drugs on the hit list may also have synergistic effects on cytokine storm, clotting disorders, secondary bacterial infections, or pulmonary function.

These drugs and natural products merit assessment in SAR-Cov-2 assays and RdRP inhibition experiments.

**Conclusions**

Our virtual screening approach which applied Autodock Vina and MDS in tandem to calculate binding energies for repurposed drugs against the RNA-dependent RNA polymerase (RdRP) identified 80 promising compounds for treating SARS-Cov2 infections. The top hits from our study consisted of a mixture of antiviral agents, natural products and drugs with diverse modes of action. The prognostic value of our computational approach was been demonstrated by our earlier studies of drugs and natural products against the viral main protease, mPro. Here we show that the same protocol generates useful predictions of agents against SARS-Cov-2 replication because a substantial number of the diverse range of drugs in the 80 compound short list exhibit useful SARS-Cov-2 antiviral effects in vitro or have been identified in other computational studies on RdRP. The antiviral drugs simeprevir, sofosbuvir, lopinavir, ritonavir and remdesivir exhibit strong antiviral properties and several in in clinical trial or use against SARS-Cov-2. These drugs have also been identified as binding to RdRP by other virtual screening studies and by in vitro assays.

Again, this high validation success rate reinforced the view that our virtual screening protocols were able to identify existing drugs and approved natural products, for rapid testing in the clinic against COVID-19. The hits may also have activity against other coronaviruses. The identified drugs may be useful for treating COVID-19 patients and provide a rational computational basis for repurposing drugs for future pandemics and other diseases.

## Materials and Methods

*Protein structure preparation and grid preparation*

The crystal structure of the SARS-CoV-2 RdRp (Figure 11) was downloaded from RCSB PDB (https://www.rcsb.org/structure/6M71) with a reported resolution of 2.90Å.

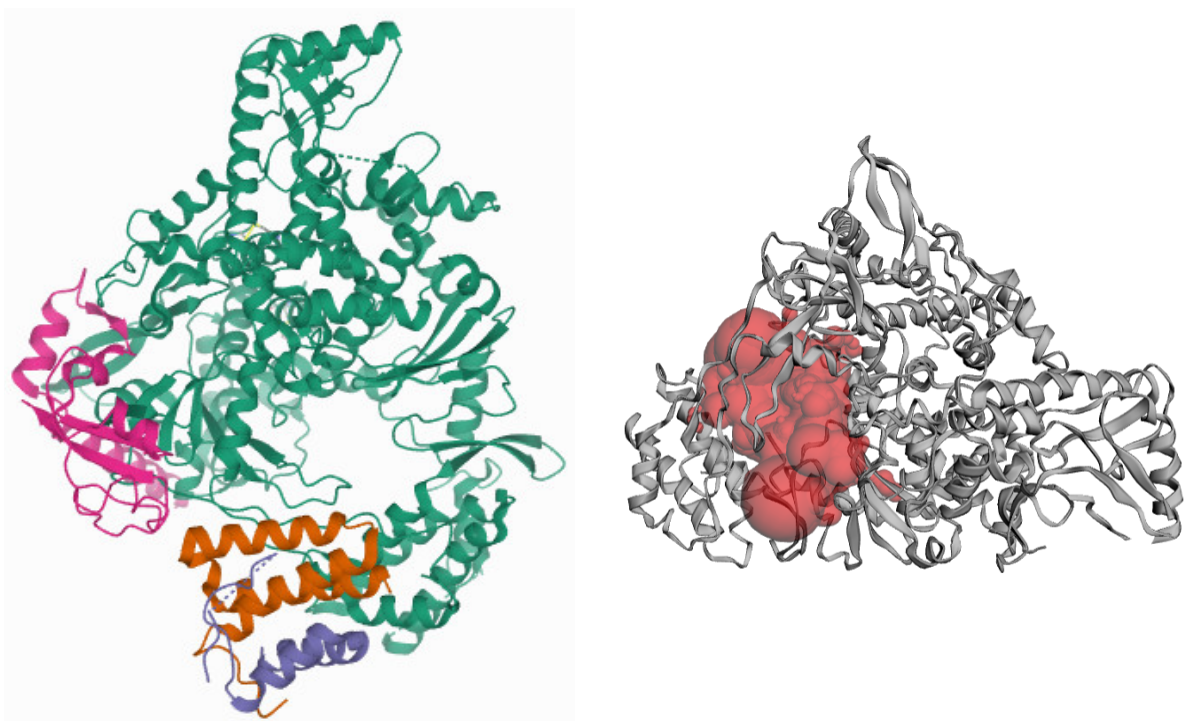

**Figure 11**. 3D structure of SARS-Cov-2 RdRP (PDB refcode 6M71) (left) and position and shape of binding pocket (right).

Protein preparation, removal of non-essential and non-bridging water molecules, addition of hydrogen atoms and missing residues and loops for docking studies were performed using UCSF Chimera package (https://www.cgl.ucsf.edu/chimera/). [75] AutoDock Tools (ADT) software was used to prepare the required files for Autodock Vina by assigning hydrogen polarities, calculating Gasteiger charges to protein structures and converting protein structures from the .pdb file format to .pdbqt format.[76].

*Screening databases*

Drugs were downloaded from the DrugBank database (Wishart et al., 2018) and CHEMBL database (FDA approved) (Gaulton et al., 2017). A total of 8773 and 13,308 drugs were retrieved from Drugbank and CHEMBL database, respectively. The drugs were dowloaded in sdf format and converted to .pdbqt format using Raccoon (Forli et al., 2016).

*Docking Methodology*

Small molecule ligand structures were docked against protein structure using the AutoDock Vina (version 1.1.3) package.[76] AutoDock Vina employs gradient-based conformational search approach and an energy-based empirical scoring function. AutoDock Vina is also flexible, easily scripted, extensively validated in many published studies with a variety of proteins and ligands and takes advantage of large multi-CPU or -GPU machines to run many calculations in parallel. The code has also been employed very successfully to dock millions of small molecule drug candidates into a series of protein targets to discover new potent drug leads. The package includes useful scripts for generating modified .pdb files required for grid calculations and for setting up the grid calculations around each protein automatically. The software requires the removal of hydrogens, addition of polar hydrogens, setting of the correct atom types, and calculation of atom charges compatible with the AutoGrid code. The algorithm generates a grid around each protein and calculates the interaction energy of a probe noble gas atom at each grid position outside and within internal cavities of the protein. The grid resolution was set to 1 Å, the maximum number of binding modes to output was fixed at 10, and the exhaustiveness level (controlling the number of independent runs performed) was set at 8. The docking employed a genetic algorithm to optimize the binding conformations of the ligands during docking to the protease site. Drugs were docked individually to the active site of RdRP (refcode 6M71) with the grid coordinates (grid centre) and grid boxes of appropriate sizes generated by the bash script vina_screen.sh (Supplementary Information). The top scored compounds were identified with a python script script1.py (Supplementary Information) and subjected to molecular dynamic simulation. The docked structures were analysed using UCSF Chimera [75] and LigPlot+ software[77] to illustrate hydrogen-bond and hydrophobic interactions. A total of fifty top compounds selected from each of the Drugbank and CHEMBL compounds. Fifteen compounds were common to both database top hits. Molecular dynamics studies were conducted on the unique set of eighty-five compounds from both sets.

*Molecular Dynamics Simulation*

The top screened compound complexes with protease were minimized with CHARMm force field. The topology files of the ligands were prepared from Swissparam (http://www.swissparam.ch/) [78] and minimized in Gromacs2020 (http://www.gromacs.org/).[79]. Docked complexes of ligands and COVID-19 M$^{pro}$ protein were used as starting geometries for MD simulations. Simulations were carried out using the GPU accelerated version of the program with the CHARMm force field I periodic boundary conditions in ORACLE server. Docked complexes were immersed in a truncated octahedron box of TIP3P water molecules. The solvated box was further neutralized with Na+ or Cl− counter ions using the tleap program. Particle Mesh Ewald (PME) was employed to calculate the long-range electrostatic interactions. The cut-off distance for the long-range van der Waals (VDW) energy term was 12.0 Å. The whole system was minimized without any restraint. The above steps applied 2500 cycles of steepest descent minimization followed by 5000 cycles of conjugate gradient minimization. After system optimization, the MD simulations was initiated by gradually heating each system in the NVT ensemble from 0 to 300 K for 50 ps using a Langevin thermostat with a coupling coefficient of 1.0/ps and with a force constant of 2.0 kcal/mol·Å2 on the complex. Finally, a production run of 20 ns of MD simulation was performed under a constant temperature of 300 K in the NPT ensemble with periodic boundary conditions for each system. During the MD procedure, the SHAKE algorithm was applied for the constraint of all covalent bonds involving hydrogen atoms. The time step was set to 2 fs. The structural stability of the complex was monitored by the RMSD and RMSF values of the backbone atoms of the entire protein. Calculations were also performed for up to 100 ns on few compounds to ensure that 20ns is sufficiently long for convergence. Duplicate production runs starting with different random seeds were also run to allow estimates of binding energy uncertainties to be determined.

The binding free energies of the protein-protein complexes were evaluated in two ways. The traditional method is to calculate the energies of solvated SARS-Cov-2 protease and small molecule ligands and that of the bound complex and derive the binding energy by subtraction.

$$\Delta G \text{ (binding, aq)} = \Delta G \text{ (complex, aq)} - (\Delta G \text{ (protein, aq)} + \Delta G \text{ (ligand, aq)}) \qquad (1)$$

We also calculated binding energies using the molecular mechanics Poisson Boltzmann surface area (MM/PBSA) tool in GROMACS that is derived from the nonbonded interaction energies of the complex. The method is also widely used method for binding free energy calculations.

MMPBSA calculations were conducted by GMXPBSA 2.1[80] a suite based on Bash/Perl scripts for streamlining MM/PBSA calculations on structural ensembles derived from GROMACS trajectories and to automatically calculate binding free energies for protein–protein or ligand–protein. GMXPBSA 2.1 calculates diverse MM/PBSA energy contributions from molecular mechanics (MM) and electrostatic contribution to solvation (PB) and non-polar contribution to solvation (SA). This tool combines the capability of MD simulations (GROMACS) and the Poisson–Boltzmann equation (APBS) for calculating solvation energy (Baker et., 2001). The g_mmpbsa tool in GROMACS was used after molecular dynamics simulations, the output files obtained were used to post-process binding free energies by the single-trajectory MMPBSA method. In the current study we considered 100 frames at equal distance from 20ns trajectory files.

Specifically, for a non-covalent binding interaction in the aqueous phase the binding free energy, $\Delta G$ (bind,aq), is: –

$$\Delta G \text{ (bind,aqu)} = \Delta G \text{ (bind,vac)} + \Delta G \text{ (bind,solv)} \qquad (2)$$

where $\Delta G$ (bind,vac) is the binding free energy in vacuum, and $\Delta G$(bind,solv) is the solvation free energy change upon binding: –

$$\Delta G \text{ (bind,solv)} = \Delta G \text{ (R:L, solv)} - \Delta G \text{ (R,solv)} - \Delta G \text{ (L,solv)} \qquad (3)$$

where $\Delta G$ (R:L,solv), $\Delta G$ (R,solv) and $\Delta G$ (L,solv) are solvation free energies of complex, receptor and ligand, respectively.

*Note added in proof*

As we noted in a previous SARS-Cov-2 drug repurposing paper targeting the main protease,[20] Guterres and Im showed how substantial improvement in protein-ligand docking results could be achieved using high-throughput MD simulations.[23] Over 56 protein targets (of 7 different protein classes) and 560 ligands this showed a 22% improvement in the area under receiver operating characteristics curve, from an initial value of 0.68 using AutoDock Vina alone to a final value of 0.83 when the Vina results were refined by MD.

**ACKNOWLEDGEMENTS**

We would also like to thank Oracle for providing their Cloud computing resources for the modelling studies described herein and in particular, Peter Winn, Dennis Ward, and Alison Derbenwick-Miller in facilitating these studies. NP, SP and PS are supported by National Institutes of Health Contract HHSN272201400053C. The content is solely the responsibility of the authors and does not necessarily represent the official views of the National Institutes of Health.

**Author contributions**

NP conceived project, analysed data, contributed to manuscript, SP and PS performed the computations, analysed data, contributed to the manuscript and DW analysed the data and contributed to manuscript.

# Computational screening of repurposed drugs and natural products against SARS-Cov-2 RdRP as potential COVID-19 therapies

Sakshi Piplani[1-2], Puneet Singh[1-2], Nikolai Petrovsky[1-2], David A. Winkler[3-6]

[1] College of Medicine and Public Health, Flinders University, Bedford Park 5046, Australia

[2] Vaxine Pty Ltd, 11 Walkley Avenue, Warradale 5046, Australia

[3] La Trobe University, Kingsbury Drive, Bundoora 3042, Australia

[4] Monash Institute of Pharmaceutical Sciences, Monash University, Parkville 3052, Australia

[5] School of Pharmacy, University of Nottingham, Nottingham NG7 2RD. UK

[6] CSIRO Data61, Pullenvale 4069, Australia

# Supplementary information

Table S1. Binding energies and published SARS-Cov-2 data for 80 top ranked small molecule ligands

| | ChemBl (C)) or Drugbank (D) ID | Name | $\Delta G_{MMPBSA}$ kcal/mol | SARS-Cov-2 data |
|---|---|---|---|---|
| 1 | C 3391662 | Paritaprevir | -54.3 | Predicted inhibitor of Mpro and RdRP.[1, 2] |
| 2 | C 1200633 | Ivermectin | -54.1 | $IC_{50}$ of 2.2 - 2.8 μM in monkey kidney cells.[3-5] |
| 3 | C 3126842 | Beclabuvir | -53.6 | |
| 4 | C 3809489 | Bemcentinib | -46.2 | 10-40% protection at 50μM in Vero cells.[6] $IC_{50}$ of 100nM and $CC_{50}$ of 4.7μM in human Huh7.5 cells and an $IC_{50}$ of 470nM and $CC_{50}$ was 1.6μM in Vero cells,[7] investigational treatment for COVID-19 (www.clinicaltrialsregister.eu), predicted to bind to Mpro.[2] |
| 5 | D 14761 | Remdesivir | -44.7 | In clinical trial for COVD-19, results equivocal,[8, 9] many in vitro reports e.g. $IC_{50}$ of 1μM and $CC_{50}$ of 275 μM in Vero cells,[10] IC50 of 11.4 μM in Vero cells and $IC_{50}$ of 1.3 μM in Calu-3 human lung cells,[11] SARS-Cov-2 RdRP $EC_{50}$ = 0.007 μM with $IC_{50}$ of 1.7 μM (Vero), 0.3 μM (Calu-3) and 0.01 μM (human airway epithelial cells),[12] computational prediction of RdRP inhibition,[13, 14] and Mpro inhibition.[2] |
| 6 | C 1751 | Digoxin | -41.2 | Predicted RdRP inhibitor,[15] $IC_{50}$ = 0.043 μM and $CC_{50}$ >10μM in Vero cells,[16] |
| 7 | D 09298 | Silibinin | -40.3 | Predicted RdRP inhibitor,[17, 18] |
| 8 | C 1236524 | Galidesvir | -40.3 | Clinical trials for COVID-19 and RdRP inhibitor,[19] predicted Mpro inhibitor,[20] |
| 9 | C 1076263 | Setrobuvir | -40.0 | Predicted SARS-Cov-2 RdRP[13, 21] and Mpro inhibitor,[22] |
| 10 | C 3707372 | Voxilaprevir | -39.5 | Experimental $EC_{50}$ >10 μM and $CC_{50}$ 16 μM in A549-hACE2 cells,[23] predicted RdRP[24] and Mpro inhibitor.[25] |
| 11 | C 2013174 | Vedroprevir | -38.5 | Predicted Mpro[26] and RdRP inhibitor,[27] |
| 12 | C 1241348 | Faldaprevir | -38.1 | Predicted Mpro,[28] PLpro,[29] and RdRP inhibitor,[27] |
| 13 | C 413 | Sirolimus (Rapamycin) | -37.5 | Clinical trial for COVID-19,[30, 31] predicted Mpro, PLpro and spike inhibition,[32, 33] |
| 14 | C 3301668 | Carbetocin | -37.4 | Predicted RdRP [34] and Mpro inhibitor,[35] |

| | ChemBl (C)) or Drugbank (D) ID | Name | $\Delta G_{MMPBSA}$ kcal/mol | SARS-Cov-2 data |
|---|---|---|---|---|
| 15 | D 01051 | Novobiocin | -37.3 | Predicted RdRP inhibition,[14] |
| 16 | C 1683590 | Eribulin | -36.6 | Predicted RdRP[36] and 2′-O-ribose methyltransferase inhibitor.[37] |
| 17 | C 442 | Ergotamine | -36.3 | Predicted $IC_{50}$ of 190µM,[38] predicted Mpro and RdRP inhibitor.[2, 39] |
| 18 | C 1957287 | Tegobuvir | -34.7 | SARS-Cov-2 $EC_{50}$ = >10 and $CC_{50}$ = 18 µM,[23] predicted RdRP and Mpro inhibitor,[40-42] |
| 19 | D 12466 | Favipiravir | -34.3 | Human trials for COVID-19,[9, 43, 44] $EC_{50}$ = 62 µM and $CC_{50}$ =400 µM (Vero cells),[45] predicted RdRP, helicase and Mpro inhibitor.[21, 46, 47] |
| 20 | D 14850 | Deleobuvir | -34.2 | Predicted Mpro [42] inhibitor. |
| 21 | C 297884 | Ciluprevir | -34.1 | Predicted Mpro [48] inhibitor, SARS-Cov-2 Mpro $IC_{50}$ = 21 µM,[49] |
| 22 | C 1200649 | Quinupristin | -33.8 | Predicted RdRP inhibitor [33] |
| 23 | D 01764 | Dalfopristin | -33.7 | Predicted RdRP inhibitor [33] |
| 24 | D 04703 | Hesperidin | -33.4 | Predicted Mpro [50-52] and RdRP inhibitor [50] |
| 25 | C 3545363 | Glecaprevir | -32.9 | Predicted Mpro [53] and RdRP inhibitor [24] |
| 26 | D 00224 | Indinavir | -32.6 | Predicted Mpro and RdRP inhibitor, [24] SARS-Cov-2 $EC_{50}$ >10 µM and $CC_{50}$ >50 µM (A549-hACE2 cells) [23] In vitro $EC_{50}$: 59 µM; $CC_{50}$ >81 µM (Vero cells),[54] |
| 27 | D 06290 | Simeprevir | -31.7 | In vitro SARS-Cov-2 $EC_{50}$ = 4 µM and $CC_{50}$ 19 µM (Vero cells), in Vitro Mpro inhibition $IC_{50}$ = 10 µM, negligible PLpro and RdRP enzyme inhibition,[55] predicted Mpro and RdRP inhibitor.[2, 56] |
| 28 | C 461101 | Eltrombopag | -31.6 | Predicted Mpro [2, 39] and RdRP inhibitor,[39, 40] in vitro SARS-Cov-2 $IC_{50}$ = 8 µM and $CC_{50}$ >50 µM (Vero cells),[57] $IC_{50}$ = 8 µM (Vero and Calu-3 cells) [11] |
| 29 | D 01126 | Dutasteride | -31.6 | Predicted Mpro and RdRP inhibitor and regulates expression of transmembrane serine protease 2 (TMPRSS2),[39] |
| 30 | C 3039514 | Elbasvir | -31.6 | SARS-Cov-2 in vitro $EC_{50}$ = 23 µM (Huh7-hACE2 cells),[58] predicted Mpro [2] and RdRP inhibition.[59] |
| 31 | C 44657 | Etoposide | -30.4 | Predicted Mpro [60, 61] |
| 32 | D 06638 | Quarfloxin | -30.2 | Predicted Mpro, human ACE2,[2, 62] and PLpro inhibitor.[63] |
| 33 | D 00445 | Epirubicin | -29.9 | Predicted Mpro,[39] |
| 34 | C 3833385 | Ruzasvir | -28.9 | Predicted Mpro, RdRP inhibitor[1, 2] |

|    | ChemBl (C)) or Drugbank (D) ID | Name | $\Delta G_{MMPBSA}$ kcal/mol | SARS-Cov-2 data |
|----|----|----|----|----|
| 35 | D 11753 | Rifamycin | -28.5 | Predicted Mpro and RdRP inhibitor.[2, 64] |
| 36 | C 1017 | Telmisartan | -28.0 | Predicted human ACE2 blocker and clinical trial for COVID-19 treatment (NCT04356495).[65] predicted Mpro and RdRP inhibitor.[66, 67] |
| 37 | D 03523 | Brequinar | -28.0 | Experimental SARS-Cov-2 $EC_{50}$ = 0.3 μM, $CC_{50}$ > 50 μM (Vero E6 cells),[68] |
| 38 | D 00619 | Imatinib | -27.7 | Experimental 80% reduction in Mpro activity at 10 μM, $EC_{50}$ = 8 μM (A549 cells),[69] predicted Mpro inhibitor,[70] experimental SARS-Cov-2 $IC_{50}$ = 3-5μM, $CC_{50}$ >30 μM,[71] |
| 39 | D 00872 | Conivaptan | -27.6 | Predicted Mpro and RdRP inhibitor,[39, 72] experimental SARS-COV-2 $IC_{50}$ = 10 μM (Vero cells),[73] $EC_{50}$ = 4μM (ACE2-A549 cells), SARS-Cov-2 Mpro $IC_{50}$ ~ 10μM (293T cells),[69] |
| 40 | D 08901 | Ponatinib | -27.1 | Experimental SARS-Cov-2 $EC_{50}$ = 1 μM, $CC_{50}$ 9 μM (HEK-293T cells),[58] predicted SARS-Cov-2 spike inhibitor.[74] |
| 41 | C 1660 | Rifapentine | -27.1 | Predicted RdRP[64, 75] and Mpro inhibitor.[76] |
| 42 | D 06144 | Sertindole | -26.6 | Predicted Mpro inhibition, but negligible experimental SARS-Cov-2 Mpro inhibition.[77] |
| 43 | C 2063090 | Grazoprevir | -26.5 | Experimental SARS-CoV-2 inhibition with $EC_{50}$ = 16 μM ($CC_{50}$ value >100 μM, Vero E6 cells),[58] predicted Mpro[78] and RdRP inhibitor.[24, 40] |
| 44 | D 01167 | Itraconazole | -26.4 | Predicted Mpro[51] and RdRP inhibition,[15] experimental SARS-Cov-2 Mpro $EC_{50}$ = 110 μM.[77] and SARS-Cov-2 $EC_{50}$ = 2.3 μM (human Caco-2 cells).[79] |
| 45 | C 2103975 | Vapreotide | -26.2 | Predicted Mpro[80] inhibition |
| 46 | D 06733 | Bafilomycin A1 | -25.9 | Experimental SARS-Cov-2 $EC_{50}$ > 50μM and $CC_{50}$ > 50 μM,[68] |
| 47 | C 1738757 | Rebastinib | -25.7 | Predicted Mpro[2] and 2'-O-ribose methyltransferase[81, 82] inhibitor. |
| 48 | D 06448 | Lonafarnib | -25.3 | Predicted RdRP[40] PLpro and Mpro[25, 63] inhibitor. |
| 49 | C 493 | Bromocriptine | -25.3 | Predicted Mrpo[2] and NSP14 inhibitor.[18] |
| 50 | D 01177 | Idarubicin | -25.2 | Predicted RdRP and Mpro inhibitor.[66, 83] Weak in vitro Mpro activity $IC_{50}$ = 250−600μM.[84] |
| 51 | C 608533 | Midostaurin | -25.0 | Predicted Mpro,[2] and spike inhibitor.[85] |
| 52 | D 01698 | Rutin | -25.0 | Predicted RdRP and Mpro inhibitor.[86] |
| 53 | C 2103882 | Tivantinib | -24.8 | Predicted Mpro and 2 2'-O-methyltransferase inhibitor.[42, 82] |

|    | ChemBl (C)) or Drugbank (D) ID | Name | $\Delta G_{MMPBSA}$ kcal/mol | SARS-Cov-2 data |
|----|---|---|---|---|
| 54 | C 490672 | Filibuvir | -24.6 | Predicted RdRP[40] and Mpro inhibitor.[42] |
| 55 | C 1983268 | Entrectinib | -23.8 | Predicted SARS-Cov-2 2'-O-methyltransferase inhibitor.[81] |
| 56 | C 1429 | Desmopressin | -23.7 | Predicted SARS-Cov-2 spike,[87] helicase,[61] spike,[88] and 2'- O-methyltransferase inhibitor.[89] |
| 57 | C 2106409 | Elsamitrucin | -22.6 | Predicted Mpro inhibitor.[90] |
| 58 | D 11618 | Zorubicin | -22.3 | Predicted inhibitor of SARS-Cov-2 spike. [91] |
| 59 | C 3039525 | Golvatinib | -22.2 | Predicted SARS-Cov-2 RDRP NSP12-NSP7 interface,[40] NSP14 SAM-dependent N7-methyl transferase,[18] Mpro,[2] and 2'-O-Ribose Methyltransferase Nsp16[82] |
| 60 | C 532 | Erythromycin | -21.6 | Predicted SARS-Cov-2 Mpro inhibitor.[92] |
| 61 | D 01092 | Ouabain | -21.3 | Predicted SARS-Cov-2 RdRP and Mpro inhibitor.[15, 42] |
| 62 | C 3318007 | Pimodivir | -21.2 | Predicted SARS-Cov-2 Mpro inhibitor. [42, 93] |
| 63 | D 04785 | Streptolydigin | -20.8 | Predicted SARS-Cov-2 RdRP inhibition.[64] |
| 64 | D 11616 | Pirarubicin | -20.3 | … |
| 65 | C 2104415 | Moxidectin | -20.2 | In vitro $IC_{50}$ = 3μM in LLC-MK2 cells,[73] predicted inhibitor of Mpro, RdRP and human ACE2 receptor.[2, 94] |
| 66 | C 3137309 | Venetoclax | -20.1 | Inactive in SARS-CoV-2 CPE assay,[95] predicted Mpro inhibitor[60] |
| 67 | D 00549 | Zafirlukast | -20.0 | Predicted inhibitor of SARS-Cov-2 Mpro, spike, and 2'-O-methyltransferases.[2, 37, 91] |
| 68 | C 4093031 | BMS-929075 | -19.8 | … |
| 69 | D 01267 | Paliperidone | -19.4 | Predicted inhibitor of SARS-Cov-2 RdRP and 2'-O-methyltransferases.[39, 82] |
| 70 | C 1951095 | Eravacycline | -18.8 | Predicted inhibitor of SARS-CoV-2 Mpro [2, 96] RdRP (NSP12), and human ACE2 receptor.[94] |
| 71 | C 372795 | Streptomycin | -18.4 | … |
| 72 | D 09280 | Lumacaftor | -17.8 | Predicted to be a SARS-Cov-2 spike protein,[85] Mpro[97] and helicase inhibitor.[98] |
| 73 | C 3989904 | Cethromycin | -17.0 | … |
| 74 | D 01254 | Dasatinib | -16.9 | Active against SARS and MERS at low μM in vitro.[99] Predicted to bind to Mpro. [100] |
| 75 | D 01410 | Ciclesonide | -16.7 | In vitro $EC_{90}$ for SARS-CoV-2 of 5μM in Vero cells and 0.55μM in differentiated human bronchial tracheal epithelial cells, blocks viral RNA replication, supresses replication of 15 mutants by >90% [57, 101] Used to treat COVID-19 patients. [102] |

|    | ChemBl (C)) or Drugbank (D) ID | Name | $\Delta G_{MMPBSA}$ kcal/mol | SARS-Cov-2 data |
|----|---|---|---|---|
| 76 | C 4297453 | Zalypsis | -16.0 | … |
| 77 | C 444172 | Zosuquidar | -15.7 | Predicted to target 2'-O-ribose methyltransferase Nsp16 of SARS-CoV-2 |
| 78 | C 1471 | Aprepitant | -15.6 | Predicted to bind to SARS-Cov-2 Mpro [80] |
| 79 | D 03325 | Tyrosyladenylate | -13.6 | Predicted to bind to SARS-CoV-2 Nsp16 2'-O-MTase [103] |
| 80 | C 408 | Troglitazone | -13.2 | Predicted to bind to SARS-Cov-2 spike[91] |

**Table S2**. Binding interactions with RdRP binding site for top 10 ranked drugs.

| No | Drug | Interacting residues |
|---|---|---|
| 1 | Beclabuvir | Arg553, Lys621, Asp618, Tyr619, Pro620, Asp623, Arg624, Ser759, Asp760, Asp761, Lys798, Glu811, Phe812, Ser814, |
| 2 | Bemcentinib | Tyr455, Arg553, Lys621, Cys622, Asp623, Asp760, Asp761, Lys798, Glu811, Ser814 |
| 3 | Digoxin | Tyr455, Arg553, Trp617, Asp618, Lys621, Cys622, Asp623, Arg624, Asp760, Asp761, Trp800, Glu811, Phe812, Cys813, Ser814. |
| 4 | Galidesvir | Asp618, Tyr619, Asp760, Asp761, Trp800, Glu811, Phe812, Ser814 |
| 5 | Ivermectin | Arg553, Arg555, Trp617, Asp618, Asp623, Ser682, Thr687, Asp760, Asp761, Glu811, Cys813, Ser814 |
| 6 | Paritaprivir | Tyr455, Arg553, Asp618, Tyr619, Pro620, Lys621, Asp623, Arg624, Asp760, Asp761, Lys798, Ser814 |
| 7 | Remdesivir | Tyr455, Arg553, Tyr619, Trp617, Lys621, Asp623, Cys622, Arg624, Asp760, Asp761, Lys798, Trp800, Glu811 |
| 8 | Setrobuvir | Arg553, Trp617, Asp618, Lys621, Pro620, Asp623, Asp760, Asp761, Glu811 |
| 9 | Silibinin | Arg553, Trp617, Asp618, Tyr619, Asp623, Thr680, Thr687, Asp760, Asp761 |
| 10 | Voxilaprevir | Lys551, Arg553, Arg555, Asp618, Tyr619, Lys621, Asp623, Asp760, Asp761, Lys798, Glu811, Cys813, Ser814 |

**Scripts**:

**1)Conf.txt**

receptor = 6Y2F.pdbqt

center_x=  9.245

center_y=  -0.788

center_z = 18.371

size_x = 50

size_y = 50

size_z = 50

num_modes = 10

exhaustiveness = 50

**2)vina_screen.sh**

```bash
#! /bin/bash
for f in CHEMBL*.pdbqt; do
   b=`basename $f .pdbqt`
   echo Processing ligand $b
mkdir -p $b
   vina --config conf.txt --cpu 50 --ligand $f --out $[b]/out.pdbqt --log $[b]/log.txt
done
```

**3)Script1.py**

```python
#! /usr/bin/env python
import sys
import glob
def doit(n):
    file_names = glob.glob('*/*.pdbqt')
    everything = []
    failures = []
    print 'Found', len(file_names), 'pdbqt files'
    for file_name in file_names:
        file = open(file_name)
        lines = file.readlines()
        file.close()
        try:
            line = lines[1]
            result = float(line.split(':')[1].split()[0])
            everything.append([result, file_name])
        except:
            failures.append(file_name)
    everything.sort(lambda x,y: cmp(x[0], y[0]))
    part = everything[:n]
    for p in part:
        print p[1],
    print
```

```
    if len(failures) > 0:
        print 'WARNING:', len(failures), 'pdbqt files could not be processed'
if __name__ == '__main__':
    doit(int(sys.argv[1]))
```